\documentclass[11pt]{article}
\usepackage{moriond}
\usepackage{float}
\usepackage{floatflt}
\usepackage{subfigure}

\usepackage[centertags]{amsmath}
\allowdisplaybreaks[1]
\numberwithin{equation}{section}
\usepackage{amsbsy}
\usepackage{amsfonts}
\usepackage{amssymb}
\usepackage[dvips]{graphicx}
\usepackage{url}
\begin{document}
\vspace*{4cm}
\title{Status of Neutrino Fits}

\author{C. Giunti}

\address{INFN, Sezione di Torino, and Dipartimento di Fisica Teorica,
\\
Universit\`a di Torino,
Via P. Giuria 1, I--10125 Torino, Italy}

\maketitle

\abstracts{
I review
the current status of
solar, KamLAND, atmospheric and K2K neutrino experiments,
implications for three-neutrino mixing
and upper bounds for the neutrino masses.
}

\raisebox{3.0cm}[0pt][0pt]{
\begin{minipage}[t]{0.905\textwidth}
\begin{center}
\includegraphics*[bb=14 14 252 240, width=4cm]{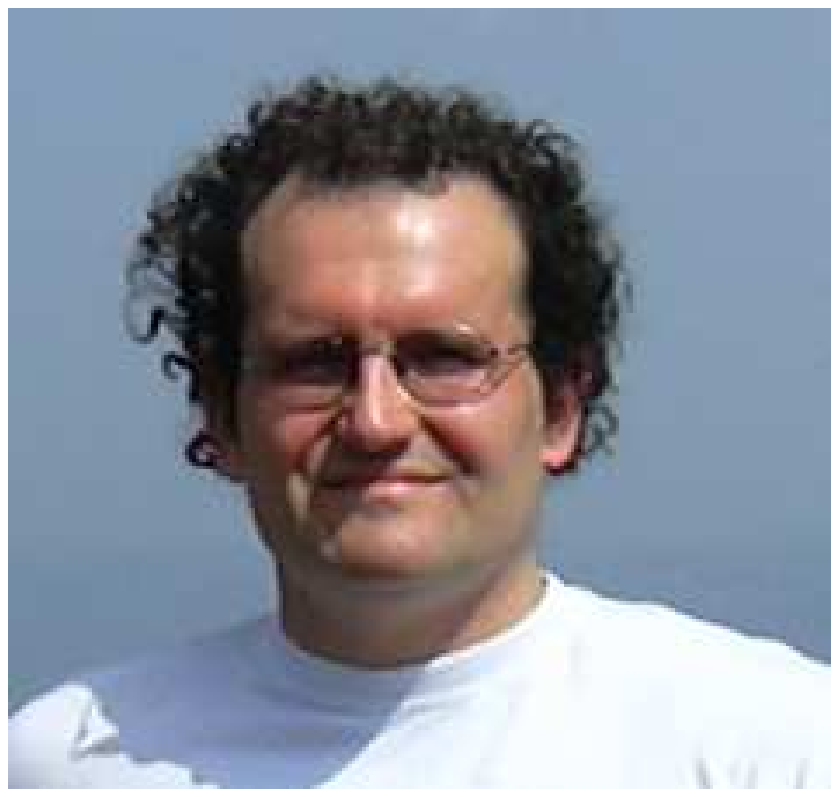}
\end{center}
\end{minipage}
}

\null
\vspace{-2.5cm}
\null

\section{Introduction}
\label{Introduction}

The last five years have been extraordinary for neutrino physics,
with the
Super-Kamiokande evidence of atmospheric neutrino oscillations~\cite{Fukuda:1998mi},
the SNO evidence of solar neutrino oscillations~\cite{Ahmad:2001an,Ahmad:2002jz},
the KamLAND~\cite{hep-ex/0212021} and K2K~\cite{Ahn:2002up}
confirmations of these evidences with neutrinos from controlled laboratory
sources.
As a result of the findings of these and other important experiments,
a scenario of bilarge mixing of three neutrinos has emerged as
favored by the data.

In this review I briefly summarize the results
of solar, KamLAND, atmospheric and K2K neutrino experiments
in Sections~\ref{Solar Neutrino Experiments and KamLAND}
and
\ref{Atmospheric Neutrino Experiments and K2K}
and their interpretation in terms of two-neutrino oscillations.
In Section~\ref{Three-Neutrino Mixing} I describe the three-neutrino mixing
schemes favored by neutrino oscillation data
and I discuss the open problem of the determination
of the absolute neutrino mass scale,
including the recent stringent limit~\cite{Spergel:2003cb}
from cosmological data.
Conclusions are drawn in Section~\ref{Conclusions}.

\section{Solar Neutrino Experiments and KamLAND}
\label{Solar Neutrino Experiments and KamLAND}

The Solar Neutrino Problem
consists in a deficit of electron neutrinos
arriving on Earth with respect to the
flux predicted by the Standard Solar Model~\cite{Bahcall:2002ng} (SSM).
It was discovered in the Homestake experiment~\cite{Cleveland:1998nv}
in the late 60's and later confirmed
by the Kamiokande~\cite{Fukuda:1996sz},
SAGE~\cite{Abdurashitov:2002nt},
GALLEX~\cite{Hampel:1998xg},
GNO~\cite{Altmann:2000ft},
Super-Kamiokande~\cite{Fukuda:2002pe}
and SNO~\cite{Ahmad:2002jz}
experiments.
Table~\ref{sundata} shows the main characteristics of
the rates measured in solar neutrino experiments.
One can see that all the experiments measuring
exclusively~\footnote{
The radiochemical experiments
Homestake~\cite{Cleveland:1998nv},
SAGE~\cite{Abdurashitov:2002nt},
GALLEX~\cite{Hampel:1998xg},
GNO~\cite{Altmann:2000ft}.
}
or mainly~\footnote{
The cross section of $\nu_e$ in the ES reaction in
Kamiokande,
Super-Kamiokande
and SNO
is about six times larger than the cross section of $\nu_{\mu,\tau}$.
In the discussion of solar neutrino experiments
$\nu_{\mu}$ and $\nu_{\tau}$
are considered together, because they are indistinguishable
due to the low energy that does not allow charged-current
reactions with production of $\mu$ or $\tau$.
}
electron neutrinos
observe a suppression between
about 1/3 and 1/2 of the electron neutrino flux
with respect to the BP2000~\cite{Bahcall:2000nu}
SSM prediction.

\begin{table}[t]
\begin{center}
\footnotesize
\setlength{\arraycolsep}{0pt}
\begin{tabular}{|c|c|c|c|c|c|}
\hline
Experiment
&
Reaction
&
$
\begin{array}{c}
E_{\mathrm{th}}
\\
\text{(MeV)}
\end{array}
$
&
$
\begin{array}{c}
\text{$\nu$ Flux}
\\
\text{Sensitivity}
\end{array}
$
&
$
\begin{array}{c}
\text{Operating}
\\
\text{Time}
\end{array}
$
&
$
\displaystyle
\frac
{ R^{\mathrm{exp}} }
{ R^{\mathrm{BP2000}} }
$
\\
\hline
\hline
SAGE
&
&
&
&
$1990-2001$
&
$ 0.55 \pm 0.05 $
\text{\protect{\cite{Abdurashitov:2002nt}}}
\vphantom{$\Big|$}
\\
\cline{1-1}
\cline{5-6}
GALLEX
&
\text{CC:}
$
\nu_e + {}^{71}\mathrm{Ga} \to {}^{71}\mathrm{Ge} + e^-
$
&
$ 0.233 $
&
\begin{tabular}{c}
\vphantom{$\Big|$}
$pp$,
${^7\mathrm{Be}}$,
${^8\mathrm{B}}$,
\\
\vphantom{$\Big|$}
$pep$,
$hep$,
\\
\vphantom{$\Big|$}
${^{13}\mathrm{N}}$,
${^{15}\mathrm{O}}$,
${^{17}\mathrm{F}}$
\end{tabular}
&
$1991-1997$
&
$ 0.61 \pm 0.06 $
\text{\protect{\cite{Hampel:1998xg}}}
\vphantom{$\Big|$}
\\
\cline{1-1}
\cline{5-6}
GNO
&
&
&
&
$1998-2000$
&
$ 0.51 \pm 0.08 $
\text{\protect{\cite{Altmann:2000ft}}}
\vphantom{$\Big|$}
\\
\hline
Homestake
&
\text{CC:}
$
\nu_e + ^{37}\mathrm{Cl}
\to
^{37}\mathrm{Ar} + e^-
$
&
$ 0.814 $
&
\begin{tabular}{c}
${^7\mathrm{Be}}$,
${^8\mathrm{B}}$,
\\
$pep$,
$hep$,
\\
${^{13}\mathrm{N}}$,
${^{15}\mathrm{O}}$,
${^{17}\mathrm{F}}$
\end{tabular}
&
$1970-1994$
&
$ 0.34 \pm 0.03 $
\text{\protect{\cite{Cleveland:1998nv}}}
\vphantom{$\Big|$}
\\
\hline
Kamiokande
&
&
$ 6.75 $
&
&
$
\begin{array}{c}
1987-1995
\\
\text{2079 days}
\end{array}
$
&
$ 0.55 \pm 0.08 $
\text{\protect{\cite{Fukuda:1996sz}}}
\vphantom{$\Big|$}
\\
\cline{1-1}
\cline{3-3}
\cline{5-6}
Super-Kam.
&
\raisebox{0.45cm}[0pt][0pt]
{\text{ES:} $ \nu + e^- \to \nu + e^- $}
&
$ 4.75 $
&
&
$
\begin{array}{c}
1996-2001
\\
\text{1496 days}
\end{array}
$
&
$ 0.465 \pm 0.015 $
\text{\protect{\cite{Fukuda:2002pe}}}
\vphantom{$\Big|$}
\\
\cline{1-3}
\cline{5-6}
&
\text{CC:}
$ \nu_e + d \to p + p + e^- $
&
$ 6.9 $
&
${^8\mathrm{B}}$
&
&
$ 0.35 \pm 0.02 $
\text{\protect{\cite{Ahmad:2002jz}}}
\vphantom{$\Big|$}
\\
\cline{2-3}
\cline{6-6}
SNO
&
\text{NC:}
$ \nu + d \to p + n + \nu $
&
$ 2.2 $
&
&
$
\begin{array}{c}
1999-2002
\\
\text{306.4 days}
\end{array}
$
&
$ 1.01 \pm 0.13 $
\text{\protect{\cite{Ahmad:2002jz}}}
\vphantom{$\Big|$}
\\
\cline{2-3}
\cline{6-6}
&
\text{ES:}
$ \nu + e^- \to \nu + e^- $
&
$ 5.2 $
&
&
&
$ 0.47 \pm 0.05 $
\text{\protect{\cite{Ahmad:2002jz}}}
\vphantom{$\Big|$}
\\
\hline
\end{tabular}
\end{center}
\caption{ \label{sundata}
Main characteristics of the rates measured in solar neutrino experiments.
CC = Charged Current;
NC = Neutral Current;
ES = Elastic Scattering;
$E_{\mathrm{th}}$ = Energy threshold.
The last column shows the measured rate normalized to the
BP2000~\protect\cite{Bahcall:2000nu}
Standard Solar Model (SSM) prediction.
}
\end{table}

In 2001~\cite{Ahmad:2001an} the SNO experiment
measured the solar neutrino flux through the
CC and ES reactions listed in Table~\ref{sundata}.
A comparison of the SNO CC rate and the
Super-Kamiokande ES rate
(which is compatible with the SNO ES rate,
but more precise because of higher statistics)
allowed the SNO Collaboration to claim a first $3.3\sigma$
evidence in favor of
$\nu_e\to\nu_{\mu,\tau}$
transitions in the solar neutrino flux.
The $\nu_{\mu,\tau}$ produced in these transitions
interact through the ES reaction causing an increase
of the ES rate
with respect to the CC rate,
normalized to the SSM prediction.

In 2002~\cite{Ahmad:2002jz},
the observation by the SNO experiment of a rate
compatible with the SSM prediction
through the NC reaction
marked the triumph of the SNO experiment,
allowing the SNO Collaboration to proof the
existence of
$\nu_e\to\nu_{\mu,\tau}$
transitions in the solar neutrino flux
using only its own data,
simply by the comparison of the fluxes measured
through the CC and NC reactions:
$
\Phi_{\nu_{\mu,\tau}}^{\mathrm{SNO}}
=
\Phi_{\mathrm{NC}}^{\mathrm{SNO}}
-
\Phi_{\mathrm{CC}}^{\mathrm{SNO}}
=
\left( 3.33 \pm 0.65 \right) \times 10^6 \, \mathrm{cm}^{-2} \, \mathrm{s}^{-1}
$.
The model-independent evidence in favor of
$\nu_e\to\nu_{\mu,\tau}$
transitions
is at the level of $5.1\sigma$.
Moreover,
the total flux of active neutrinos
measured in the SNO NC reaction
is perfectly compatible with
the SSM flux,
providing a strong evidence in favor of
the correctness of the SSM
(that was questioned before)
and restricting possible
transitions of solar $\nu_e$'s into sterile neutrinos.

The SNO measurements have solved the long-standing
Solar Neutrino Problem
proving that it is due to new neutrino physics,
because no astrophysical explanation can
generate the observed $\nu_{\mu,\tau}$'s.

The simplest known explanation of the appearance of
$\nu_{\mu,\tau}$
in the solar neutrino flux is neutrino oscillations
due to neutrino mixing~\footnote{
B. Pontecorvo discovered neutrino oscillations in
1958-59~\cite{Pontecorvo:1957cp,Pontecorvo-58}
and worked on the theory and phenomenology of neutrino oscillations
for many years (see the review in Ref.~\cite{Bilenky-Pontecorvo-PR-78}).
It is interesting to notice that
Pontecorvo predicted a possible
suppression of the solar electron neutrino flux in
1967~\cite{Pontecorvo-68},
before the first data of the Homestake experiment,
which was aimed at the observation of solar $^{8}\mathrm{B}$
in order to check the thermonuclear origin of the Sun
energy~\cite{Bahcall:1964gx,Davis:1964hf}.
}:
the left-handed neutrino fields
$\nu_{\alpha L}$ ($\alpha=e,\mu,\tau$),
which appear in the weak charged current
$
j^{\mathrm{CC}}_{\rho}
=
2
\sum_{\alpha=e,\mu,\tau}
\overline{\nu_{\alpha L}} \gamma_\rho \alpha
$,
are mixings of the neutrino fields $\nu_k$
with masses $m_k$ ($k=1,2,3,\ldots$)
according to the relation
$
\nu_{\alpha L}
=
\sum_k
U_{\alpha k}
\nu_{k L}
$,
where $U$ is the unitary mixing matrix.
The modification of the oscillation probability
due to neutrino propagation in the Sun and the Earth
(MSW effect~\cite{Wolfenstein:1978ue,Mikheev:1986wj})
must be taken into account.

\begin{figure}[t]
\begin{minipage}[t]{0.47\textwidth}
\includegraphics*[bb=0 0 434 360, width=0.90\textwidth]{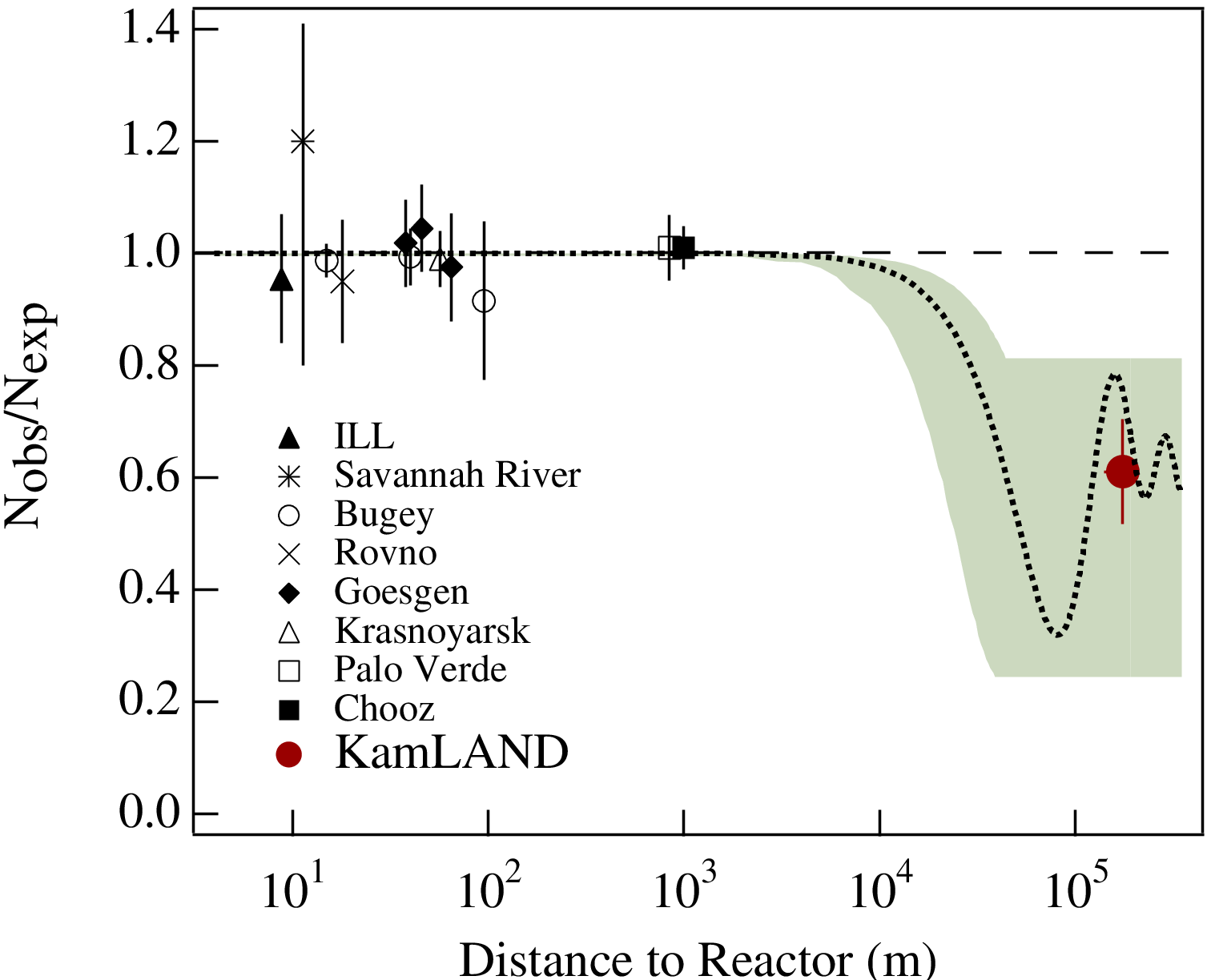}
\caption{ \label{kamland-0212021-f04}
Ratio of measured to expected $\bar\nu_e$ flux in reactor 
experiments~\protect\cite{hep-ex/0212021}.
The shaded region (dotted curve) corresponds to the 95\% C.L. LMA
region (best-fit)~\protect\cite{Fogli:2002pt}.
}
\end{minipage}
\hfill
\begin{minipage}[t]{0.47\textwidth}
\begin{center}
\includegraphics*[bb=0 0 567 539, width=0.78\textwidth]{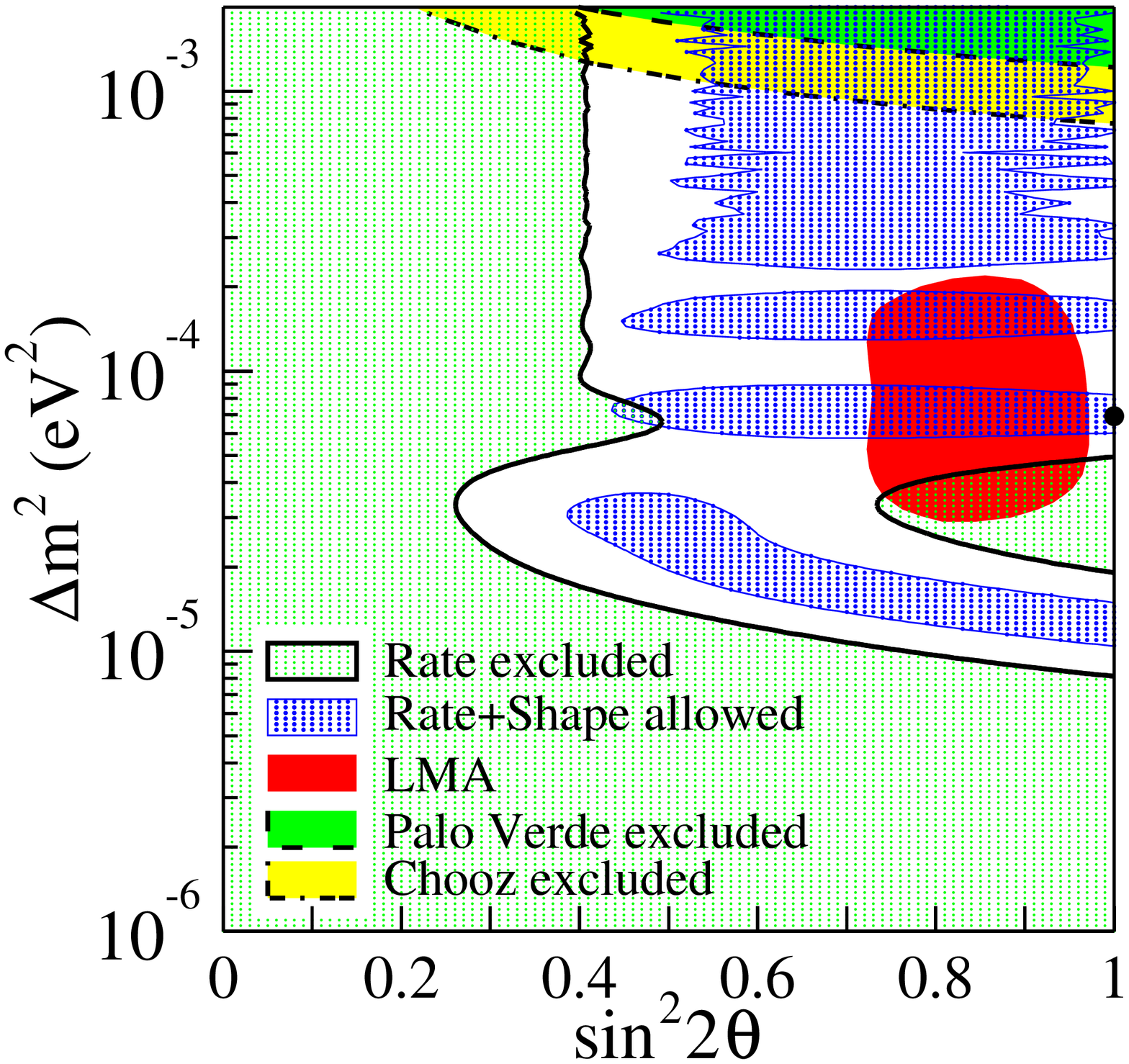}
\end{center}
\caption{ \label{kamland-0212021-f06}
KamLAND~\protect\cite{hep-ex/0212021}
excluded and allowed regions at 95\% C.L.
($\theta=\vartheta_{\mathrm{SUN}}$
and
$\Delta{m}^{2}=\Delta{m}^{2}_{\mathrm{SUN}}$).
The dark area is the 95\% C.L. LMA
region~\protect\cite{Fogli:2002pt}.
The thick dot is the best fit of
KamLAND data.
}
\end{minipage}
\end{figure}

The data of solar neutrino experiments
have been analyzed in several papers
(see Ref.~\cite{Giunti:2003tc} and references therein)
in terms of two-neutrino $\nu_e\to\nu_{\mu,\tau}$
oscillations,
leading to a rather compelling indication
in favor of the so-called LMA region
in the plane of the mixing parameters
$\tan^2\vartheta_{\mathrm{SUN}}$, $\Delta{m}^2_{\mathrm{SUN}}$
($\vartheta_{\mathrm{SUN}}$ is the effective two-neutrino mixing angle
and
$\Delta{m}^2_{\mathrm{SUN}}$ is the squared mass difference;
see Refs.~\cite{Bilenky-Pontecorvo-PR-78,BGG-review-98}),
at
$\tan^2\vartheta_{\mathrm{SUN}} \approx 0.4$
and
$\Delta{m}^2_{\mathrm{SUN}} \approx 5 \times 10^{-5} \, \mathrm{eV}^2$.

\begin{figure}[t]
\begin{minipage}[t]{0.47\textwidth}
\includegraphics*[bb=6 21 439 436, width=0.80\textwidth]{fig/maltoni-0212129-f02.eps}
\caption{ \label{maltoni-0212129-f02}
Allowed 90\%, 95\%, 99\%, 99.73\% ($3\sigma$) C.L. regions
obtained from the fit of solar and KamLAND
data~\protect\cite{hep-ph/0212129}
($\theta=\vartheta_{\mathrm{SUN}}$
and
$\Delta{m}^{2}=\Delta{m}^{2}_{\mathrm{SUN}}$).
The lines delimit the allowed regions from solar data.
}
\end{minipage}
\hfill
\begin{minipage}[t]{0.47\textwidth}
\includegraphics*[bb=18 254 278 550, width=0.68\textwidth]{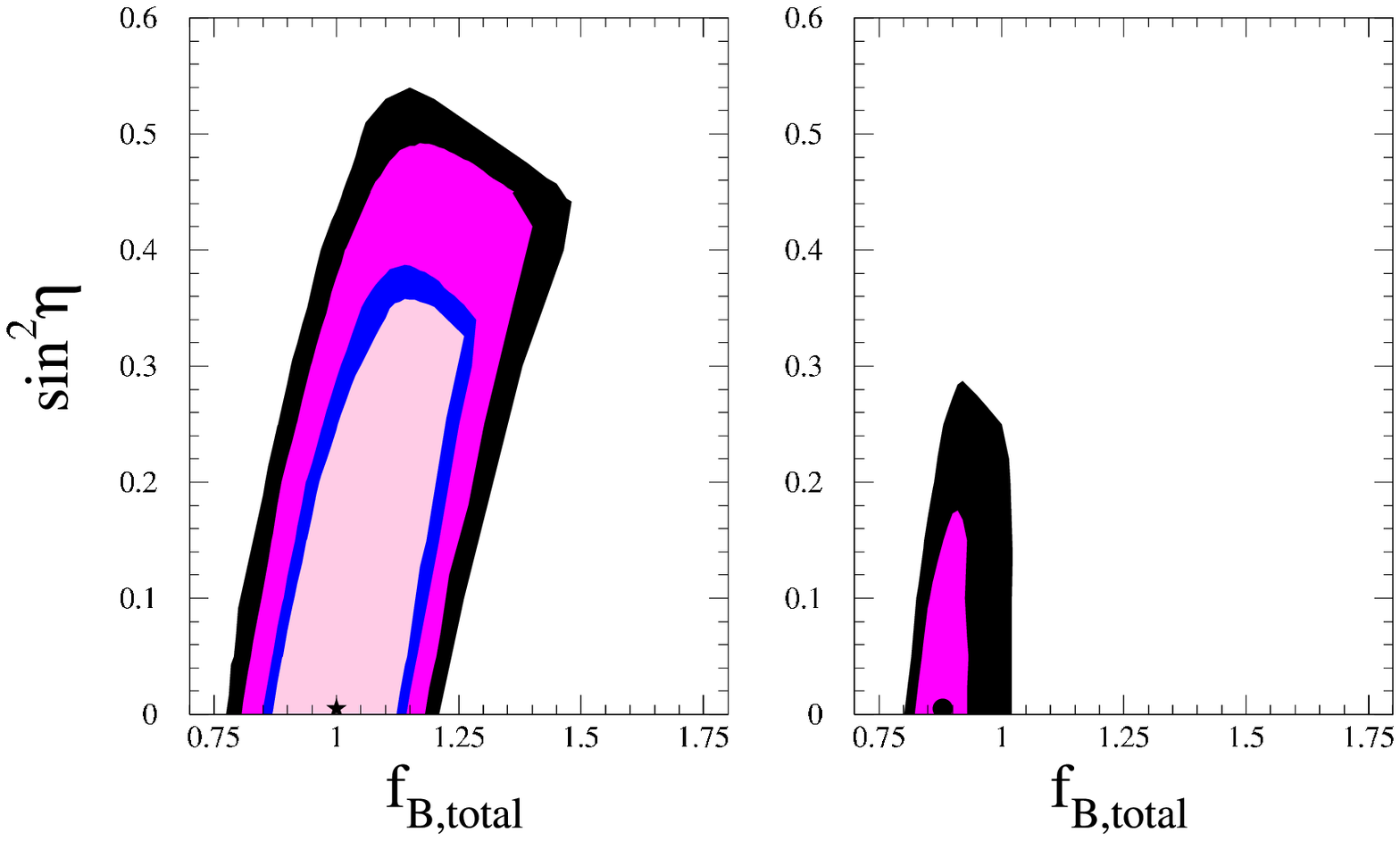}
\caption{ \label{bahcall-0212147-f06}
Allowed 90\%, 95\%, 99\%, 99.73\% ($3\sigma$) C.L. regions~\protect\cite{hep-ph/0212147}
in the $\mathrm{f_{B,total}}$--$\sin^2\eta$ plane,
where $\mathrm{f_{B,total}}=\Phi_{^8\mathrm{B}}/\Phi_{^8\mathrm{B}}^{\mathrm{SSM}}$
and
$\sin^2\eta$ is the fraction of sterile neutrinos in the solar flux.
}
\end{minipage}
\end{figure}

In December 2002 the KamLAND Collaboration
published a spectacular confirmation of the
LMA region obtained with their
long-baseline reactor $\bar\nu_e$
oscillation experiment~\cite{hep-ex/0212021}
(see also the contribution of K.~Inoue in these Proceedings).
As shown in Fig.~\ref{kamland-0212021-f04},
KamLAND is the first reactor experiment that finds
a suppression of the measured $\bar\nu_e$ flux with respect to
the original one,
thanks to the larger source-detector distance.
The measured suppression is
$
R^{\bar\nu_e}_{\mathrm{KamLAND}}
=
0.611 \pm 0.085 \pm 0.041
$.
Figure~\ref{kamland-0212021-f06}
shows the regions in the
$\sin^2 2\vartheta_{\mathrm{SUN}}$--$\Delta{m}^2_{\mathrm{SUN}}$
plane allowed by the KamLAND data.
One can see that there are two areas of overlap
with the LMA region obtained from solar neutrino data.
Therefore,
the KamLAND result is perfectly compatible with
solar neutrino data and one can expect that
a combined fit would yield two allowed subareas of the
LMA region.
Indeed,
this is what happens, as shown in
Fig.~\ref{maltoni-0212129-f02}~\cite{hep-ph/0212129}
(see Ref.~\cite{Giunti:2003tc} and references therein
for similar results obtained by other authors).
The best-fit point is~\cite{hep-ph/0212129}
\begin{equation}
\tan^2 \vartheta_{\mathrm{SUN}}
\simeq
0.46
\,,
\qquad
\Delta{m}^2_{\mathrm{SUN}}
\simeq
6.9 \times 10^{-5} \, \mathrm{eV}^2
\,,
\label{sun-bf}
\end{equation}
and
$0.29 < \tan^2\vartheta_{\mathrm{SUN}} < 0.86$
at 99.73\% C.L.

Figure~\ref{bahcall-0212147-f06}~\cite{hep-ph/0212147}
shows that the fraction of transitions of
solar $\nu_e$'s into sterile neutrinos,
given by $\sin^2\eta$,
is limited
($\sin^2\eta < 0.52$
at $3\sigma$),
and the measured total flux of solar $^8\mathrm{B}$ neutrinos
is in perfect agreement with the one predicted by the SSM,
$\Phi_{^8\mathrm{B}} = 1.00 \pm 0.06 \, \Phi_{^8\mathrm{B}}^{\mathrm{SSM}}$.

The increase of statistics of KamLAND data
will hopefully allow to distinguish among the two LMA regions
shown in Fig.~\ref{maltoni-0212129-f02}
in the near future
(see the contribution of K.~Inoue in these Proceedings).

\section{Atmospheric Neutrino Experiments and K2K}
\label{Atmospheric Neutrino Experiments and K2K}

In 1998 the Super-Kamiokande experiment discovered~\cite{Fukuda:1998mi}
an up-down asymmetry of high-energy events generated
by atmospheric $\nu_\mu$'s,
confirming the atmospheric neutrino anomaly
discovered about ten years earlier by the
Kamiokande~\cite{Hirata:1992ku}
and
IMB~\cite{Becker-Szendy:1992hq}
experiments.

Cosmic rays interacting with nuclei in the atmosphere produce pions,
which decaying produce muons and neutrinos.
The muons that decay before hitting the ground
produce more neutrinos.
These neutrinos from pion and muon decay in the atmosphere are called ``atmospheric neutrinos''.
In the late 80's the
Kamiokande
and
IMB
experiments discovered an anomaly in the ratio of events produced
by muon and electron neutrinos
with respect to the calculated one~\cite{Gaisser:2002jj},
that indicated a disappearance
of atmospheric muon neutrinos.

From simple geometry,
the fluxes of high energy neutrinos
(whose parent cosmic rays are not deflected by the magnetic field of the Earth)
coming from above and below are equal.
The $7\sigma$ Super-Kamiokande up-down asymmetry of high-energy $\nu_\mu$ events,
$
A_{\mu}
=
\left[ \left(U-D\right) / \left(U+D\right) \right]_{\mu}
=
- 0.311 \pm 0.043 \pm 0.01
$,
is a model-independent proof of the disappearance
of atmospheric muon neutrinos
generated on the other side of the Earth.

The simplest explanation of the disappearance of
atmospheric muon neutrinos
is neutrino oscillations.
The hypothesis of
$\nu_\mu \to \nu_\tau$ transitions
provides the best fit of the atmospheric neutrino data
(the $\nu_\mu \leftrightarrows \nu_e$ channel
is practically excluded
and the $\nu_\mu \to \nu_s$ channel
is strongly disfavored)~\cite{hep-ex/0206016}.
The data of the Soudan-2~\cite{Allison:1999ms} and MACRO~\cite{Ambrosio:2003yz}
atmospheric neutrino experiments
confirm this interpretation,
albeit with less precision than the Super-Kamiokande experiment.

\begin{figure}[t]
\begin{minipage}[l]{0.47\textwidth}
\includegraphics*[bb=20 145 551 646, width=\textwidth]{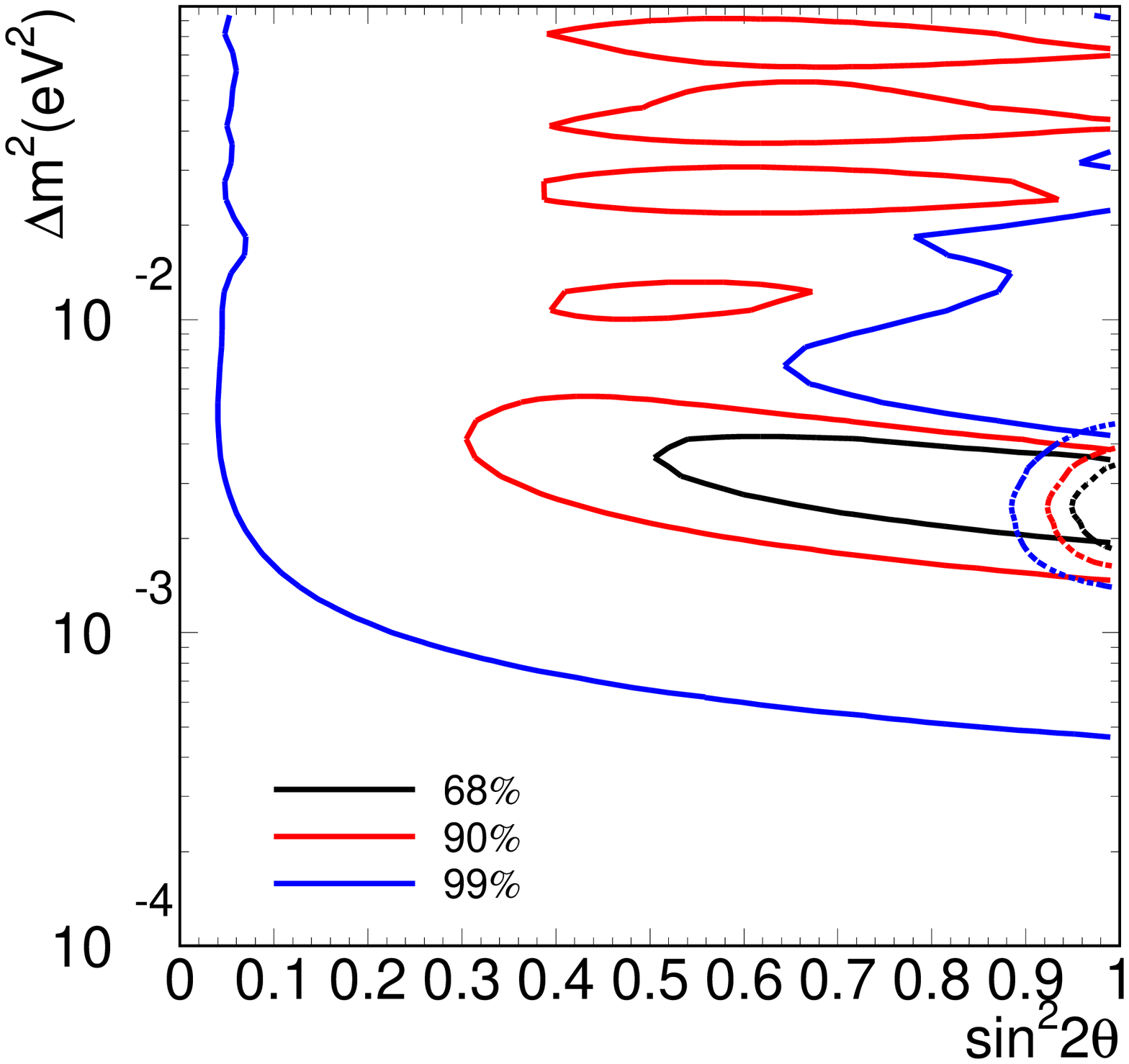}
\caption{ \label{k2k-0210030-f09}
Allowed K2K regions (large) compared with the
atmospheric Super-Kamiokande allowed regions
(small regions)~\protect\cite{hep-ex/0210030}.
$\theta=\vartheta_{\mathrm{ATM}}$
and
$\Delta{m}^{2}=\Delta{m}^{2}_{\mathrm{ATM}}$.
}
\end{minipage}
\hfill
\begin{minipage}[l]{0.47\textwidth}
\begin{center}
\includegraphics*[bb=42 299 86 509, scale=0.9]{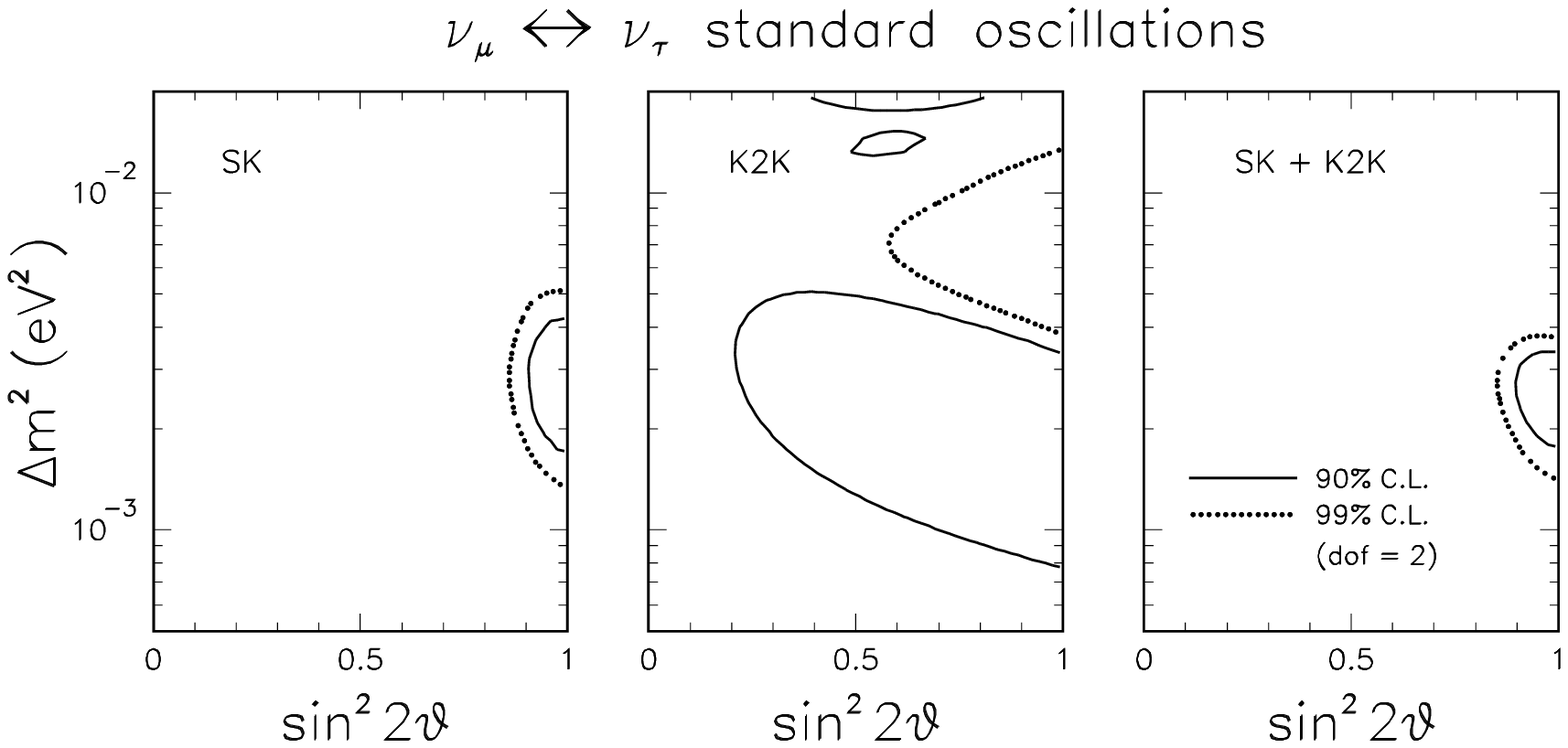}
\includegraphics*[bb=398 299 536 509, scale=0.9]{fig/fogli-0303064-f01.eps.pdf.eps}
\end{center}
\caption{ \label{fogli-0303064-f01}
Regions
allowed by a combined fit of the Super-Kamiokande atmospheric data
and K2K data~\protect\cite{hep-ph/0303064}
($\vartheta=\vartheta_{\mathrm{ATM}}$
and
$\Delta{m}^{2}=\Delta{m}^{2}_{\mathrm{ATM}}$).
}
\end{minipage}
\end{figure}

The regions in the $\sin^2 2\vartheta_{\mathrm{ATM}}$--$\Delta{m}^2_{\mathrm{ATM}}$
plane allowed by Super-Kamiokande atmospheric neutrino data
are shown in Fig.~\ref{k2k-0210030-f09}:
the small regions centered around the best-fit point
\begin{equation}
\sin^2 2\vartheta_{\mathrm{ATM}} = 1
\,,
\qquad
\Delta{m}^2_{\mathrm{ATM}} = 2.5 \times 10^{-3} \, \mathrm{eV}^2
\,.
\label{SK-atm-bf}
\end{equation}

The oscillation explanation of the disappearance of
atmospheric muon neutrinos
has been recently confirmed in
the long-baseline accelerator K2K experiment~\cite{Ahn:2002up}
in which muon neutrinos
are produced in a laboratory
by pion decay.
The K2K experiment measured 56 events instead of
$80.1 {}^{+6.2}_{-5.4}$
events expected in absence of oscillations,
with a probability smaller than 1\% that the suppression is due
to a statistical fluctuation.
The large allowed regions
in Fig.~\ref{k2k-0210030-f09},
obtained from a two-neutrino oscillation fit of the K2K data,
are in beautiful agreement
with the Super-Kamiokande allowed regions
from atmospheric data.

Figure~\ref{fogli-0303064-f01}
shows the allowed region in the
$\sin^2 2\vartheta_{\mathrm{ATM}}$--$\Delta{m}^2_{\mathrm{ATM}}$
plane
resulting from a combined fit of the Super-Kamiokande atmospheric data
and K2K data~\cite{hep-ph/0303064}.
The main effect of the K2K data is to
slightly lower the upper bound on
$\Delta{m}^2_{\mathrm{ATM}}$,
because a
$\Delta{m}^2_{\mathrm{ATM}} \sim (4-6) \times 10^{-3} \, \mathrm{eV}$
would produce an unobserved suppression of the K2K energy spectrum peak~\cite{hep-ph/0303064}.

In the near future K2K will take more data,
hopefully confirming the agreement with the Super-Kamiokande
evidence of atmospheric neutrino oscillations.
Within some years the long-baseline accelerator experiments
MINOS and CNGS
will improve the precision of the
determination of
$\sin^2 2\vartheta_{\mathrm{ATM}}$
and
$\Delta{m}^2_{\mathrm{ATM}}$
and will check the
$\nu_\mu\to\nu_\tau$
explanation
(see Ref.~\cite{hep-ex/0205043}).

\section{Three-Neutrino Mixing}
\label{Three-Neutrino Mixing}

\begin{figure}[t]
\begin{minipage}[t]{0.47\textwidth}
\includegraphics*[bb=118 426 465 753, width=\textwidth]{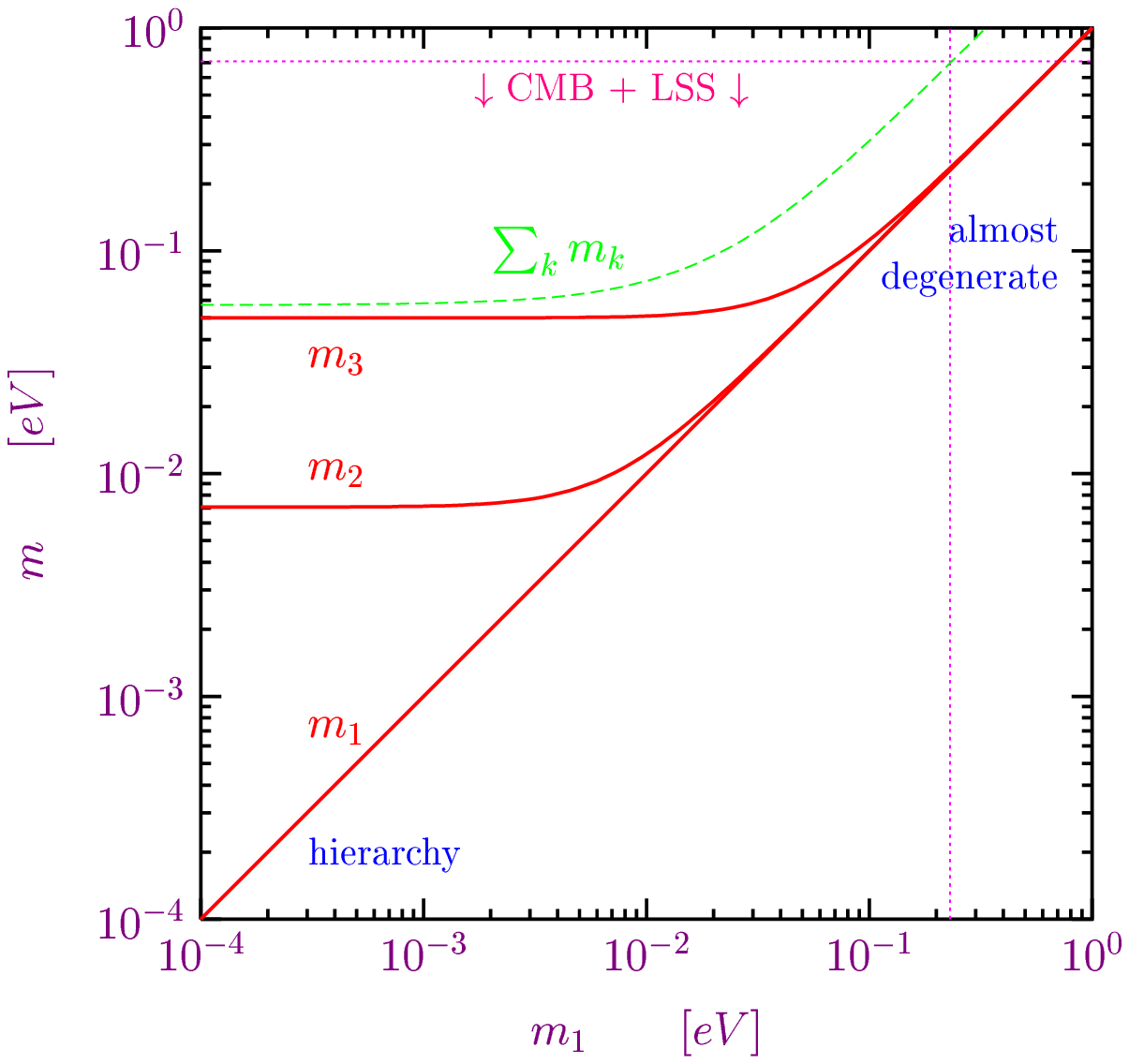}
\caption{ \label{smn}
Values of the neutrino masses as functions of the lightest mass $m_1$
in the normal scheme in Fig.~\ref{3nu}.
}
\end{minipage}
\hfill
\begin{minipage}[t]{0.47\textwidth}
\includegraphics*[bb=118 426 465 753, width=\textwidth]{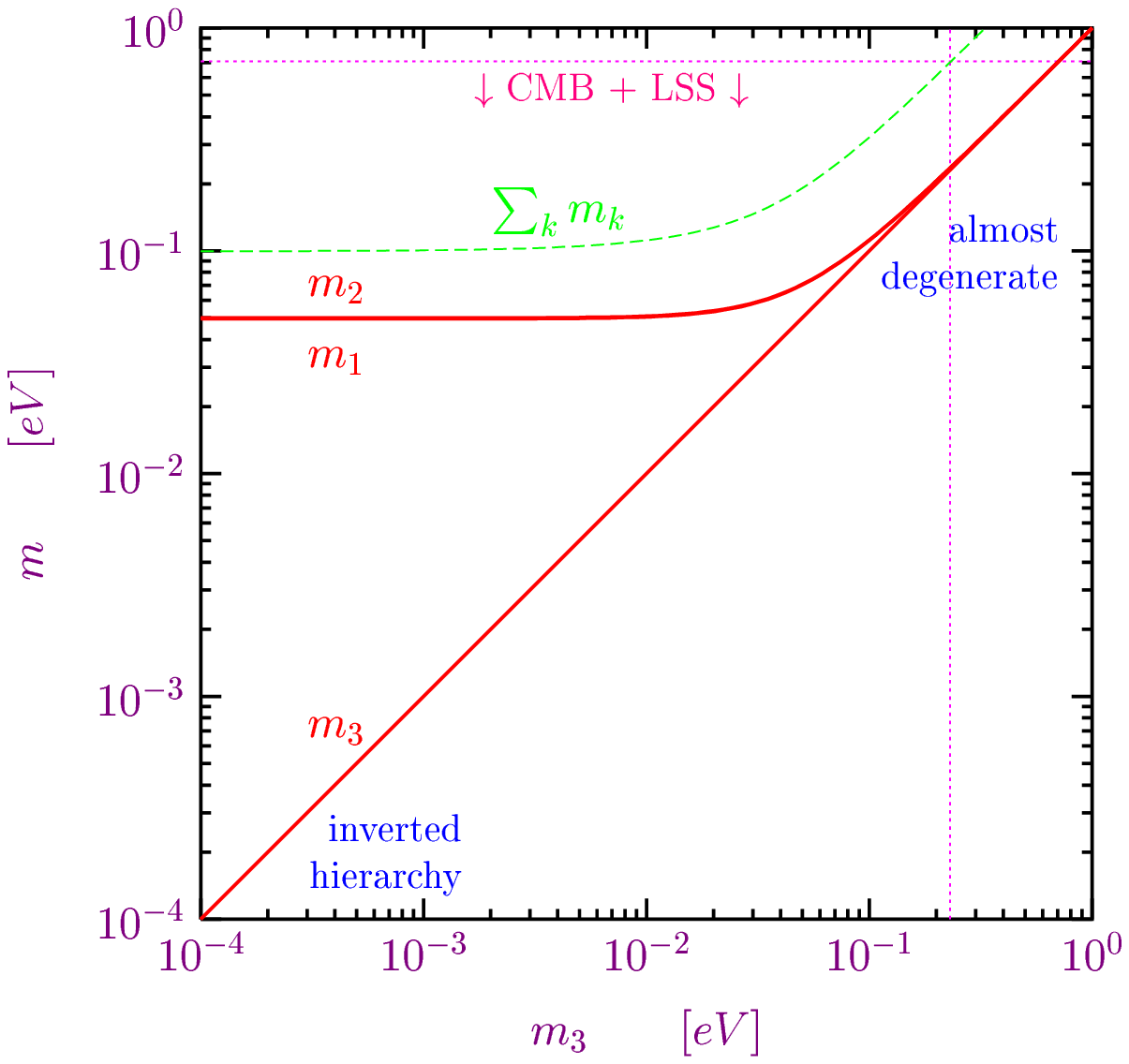}
\caption{ \label{smi}
Values of the neutrino masses as functions of the lightest mass $m_3$
in the inverted scheme in Fig.~\ref{3nu}.
}
\end{minipage}
\end{figure}

The results of solar, KamLAND, atmospheric and K2K neutrino experiments
can be explained comfortably by the hypothesis of
three-neutrino mixing,
which is the simplest and most natural one
given our knowledge of the existence of only tree active flavor neutrinos,
$\nu_e$, $\nu_\mu$ and $\nu_\tau$~\footnote{
Although the number of active flavor neutrinos is known to be three,
the number of massive neutrinos could be larger than three,
if there are sterile neutrinos
(see Ref.~\cite{BGG-review-98}).
The additional massive neutrinos may be needed
in order to generate a $\Delta{m}^2$ of the
order of 1 eV necessary to explain the
$\bar\nu_\mu\to\bar\nu_e$
transitions observed in the short-baseline LSND experiment~\cite{LSND}.
The existence of these transitions is
under investigation in the MiniBooNE experiment~\cite{hep-ex/0210020}.
}.
Figure~\ref{3nu}
shows the two three-neutrino schemes compatible with
the hierarchy of solar and atmospheric $\Delta{m}^2$'s
in Eqs.~(\ref{sun-bf}) and (\ref{SK-atm-bf}).
Since the absolute scale of neutrino masses is not fixed by oscillation data,
the two schemes allow a hierarchy
(normal or inverted)
of the neutrino masses
if one mass is much smaller than the other two,
or an almost degeneracy of the three neutrino masses,
as shown in Figs.~\ref{smn} and \ref{smi}.

The hierarchy $\Delta{m}^2_{\mathrm{SUN}} \ll \Delta{m}^2_{\mathrm{ATM}}$
implies that
solar (and KamLAND) neutrino oscillations depend only on the
first row of the mixing matrix
($U_{e1}$, $U_{e2}$, $U_{e3}$)
and
atmospheric (and K2K) neutrino oscillations depend only on the
third column of the mixing matrix
($U_{e3}$, $U_{\mu3}$, $U_{\tau3}$).
Indeed,
since $\nu_\mu$ and $\nu_\tau$ are indistinguishable
in solar neutrino experiments
(because of the low energy that allows only neutral-current interactions),
solar neutrino oscillations can be expressed in terms of the
probability of $\nu_e$ disappearance,
which depends only on the elements
$U_{e1}$, $U_{e2}$, $U_{e3}$
of the mixing matrix.
In the case of atmospheric neutrino oscillations,
since $\Delta{m}^2_{\mathrm{SUN}}=\Delta{m}^2_{21}$
is too small to have any effect,
$\nu_1$ and $\nu_2$ are indistinguishable
and all probabilities depend on
$|U_{\alpha3}|^2=1-|U_{\alpha1}|^2-|U_{\alpha2}|^2$,
for
$\alpha=e,\mu,\tau$
(in other words,
the oscillations are CP-invariant and do not depend on the mixing angle $\theta_{12}$
in the standard parameterization of the mixing matrix;
see Ref.~\cite{Giunti:1998fx}).

\begin{floatingfigure}[l]{0.43\textwidth}
\begin{flushleft}
\begin{minipage}[t]{0.43\textwidth}
\setlength{\tabcolsep}{0cm}
\begin{tabular}{lr}
\includegraphics*[bb=181 466 428 775, width=0.49\textwidth]{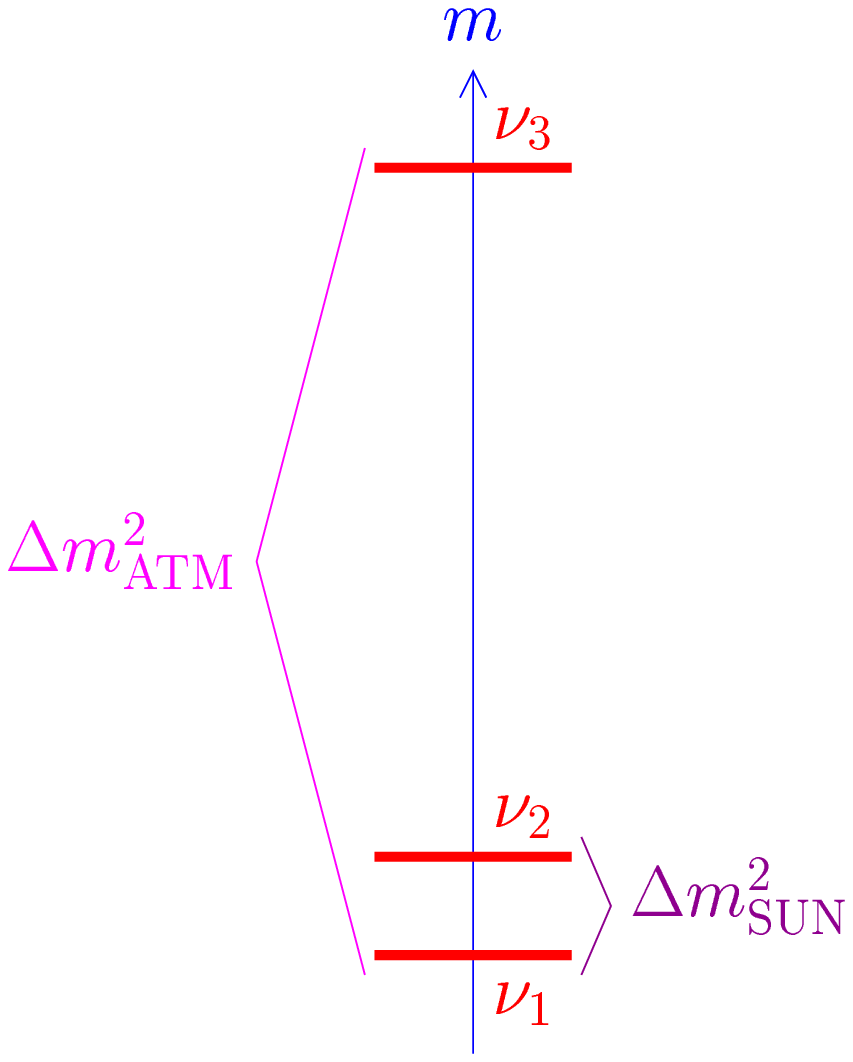}
&
\includegraphics*[bb=183 466 432 775, width=0.49\textwidth]{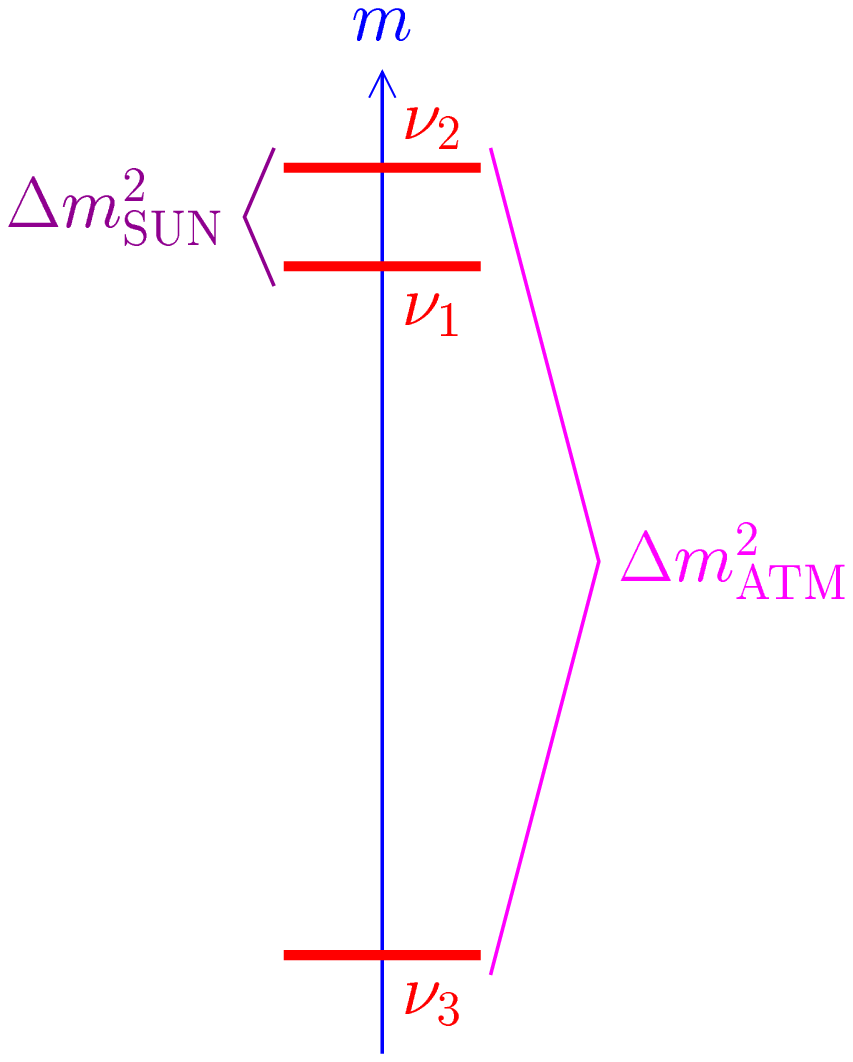}
\\
``normal''
&
``inverted''
\end{tabular}
\caption{ \label{3nu}
The two three-neutrino schemes allowed by the hierarchy
$\Delta{m}^2_{\mathrm{SUN}} \ll \Delta{m}^2_{\mathrm{ATM}}$.
The massive neutrinos are labeled in order to have
$\Delta{m}^2_{\mathrm{SUN}}=\Delta{m}^2_{21}$
and
$\Delta{m}^2_{\mathrm{ATM}}=|\Delta{m}^2_{31}|$.
}
\end{minipage}
\end{flushleft}
\end{floatingfigure}

The only element of the mixing matrix
that connects solar and atmospheric neutrino oscillations is
$U_{e3}$.
Therefore, any information on the value of
$U_{e3}$
is of crucial importance for the interpretation of
solar and atmospheric neutrino data
in the framework of three-neutrino mixing.
Such very important information comes from the
CHOOZ~\cite{Apollonio:2003gd}
and
Palo Verde~\cite{Boehm:2001ik}
long-baseline reactor experiments,
which did not observe any disappearance of $\bar\nu_e$'s
through oscillations due to $\Delta{m}^2_{\mathrm{ATM}}$,
leading to the upper bound
$
|U_{e3}|^2 < 5 \times 10^{-2}
$
(99.73\% C.L.)~\cite{Fogli:2002pb}.
This limit is very important,
because it means that solar and atmospheric neutrino oscillations
are practically decoupled effective two-neutrino oscillations~\cite{Bilenky:1998tw}.
Hence, the analyses of solar and atmospheric neutrino data
in terms of two-neutrino mixing
give information on three-neutrino mixing.

Expressing
the mixing angles measured in solar--KamLAND,
atmospheric--K2K and CHOOZ--Palo Verde experiments
in terms of the elements of the mixing matrix as
$
\sin^2 2 \vartheta_{\mathrm{SUN}} = 4 |U_{e1}|^2 |U_{e2}|^2
$,
$
\sin^2 2 \vartheta_{\mathrm{ATM}} = 4 |U_{\mu3}|^2 \left( 1 - |U_{\mu3}|^2 \right)
$,
$
\sin^2 2 \vartheta_{\mathrm{CH-PV}} = 4 |U_{e3}|^2 \left( 1 - |U_{e3}|^2 \right)
$,
the results of these experiments imply the
so-called bilarge mixing matrix~\cite{hep-ph/0212142}
\begin{equation}
|U|
\simeq
\begin{pmatrix}
0.70-0.87 & 0.50-0.69 & 0.00-0.16
\\
0.20-0.61 & 0.34-0.73 & 0.60-0.80
\\
0.21-0.63 & 0.36-0.74 & 0.58-0.80
\end{pmatrix}
\,,
\label{mixmat}
\end{equation}
in which all the elements are large, except $U_{e3}$.
In the future,
the most pressing task is to measure the value of
$|U_{e3}|$
(see the contribution of M. Mezzetto in these Proceedings),
because if $|U_{e3}|$ is not too small,
in future experiments
it may be possible to distinguish between the normal and inverted schemes in Fig.~\ref{3nu}
(through different matter effects)
and to measure CP violation in neutrino oscillations.

Another important open question is the
absolute value of neutrino masses (see Ref.~\cite{Bilenky:2002aw}).
The Mainz and Troitsk tritium $\beta$-decay experiments
provide the upper bound
$m_{\nu_e} < 2.2 \, \mathrm{eV}$
(95\% C.L.)~\cite{hep-ex/0210050}
on the effective electron neutrino mass,
which translates into an upper bound of
$2.2 \, \mathrm{eV}$ for all the three neutrino masses,
because of the small $\Delta{m}^2$'s
in Eqs.~(\ref{sun-bf}) and (\ref{SK-atm-bf}).
In the future the KATRIN experiment~\cite{hep-ex/0109033}
will explore $m_{\nu_e}$ down to about
$0.3 \, \mathrm{eV}$.

A rather tight upper bound for the
neutrino masses has been obtained very recently
through a combined analysis of cosmological data,
following the first results of the WMAP satellite experiment~\cite{Spergel:2003cb}
(see also the contribution of M.~Limon in these Proceedings).
The fit of Cosmic Microwave background (CMB) and Large Scale Structure (LSS) data
in the framework of standard Big-Bang cosmology
gives the upper bound
\begin{equation}
\sum_k m_k < 0.71 \, \mathrm{eV}
\qquad
\text{(95\% confidence)}
\,.
\label{WMAP}
\end{equation}
This limit is shown by the horizontal dotted line in
Figs.~\ref{smn} and \ref{smi},
with the sum of neutrino masses given by the dashed line.
From these figures it is clear that
the limit can be saturated only if the three neutrino masses are
almost degenerate, in both normal and inverted schemes.
Therefore,
as shown by the vertical dotted line in
Figs.~\ref{smn} and \ref{smi},
the cosmological upper limit on each neutrino mass is one third of that in Eq.~(\ref{WMAP}):
$m_k < 0.23 \, \mathrm{eV}$
(95\% confidence).
This upper bound is very impressive and appears already
stronger than the sensitivity of about
$0.3 \, \mathrm{eV}$
of the future KATRIN experiment~\cite{hep-ex/0109033}.
Let me however emphasize that the KATRIN experiment
is nevertheless very important to confirm the cosmological bound,
which follows from several plausible but not certain assumptions.
Even more interesting would be a measurement of
a neutrino mass above the cosmological bound,
which would demand a non-standard explanation.

\section{Conclusions}
\label{Conclusions}

We have seen that the last years have been exceptionally fruitful
for neutrino physics.
Important experimental discoveries
give us a rather convincing picture of neutrino mixing.
However,
there are still important questions that need to be clarified:
\begin{enumerate}
\setlength{\topsep}{0cm}
\setlength{\partopsep}{0cm}
\setlength{\parsep}{0cm}
\setlength{\itemsep}{0cm}
\item
Which is the absolute scale of neutrino masses?
At present we have only upper bounds from direct measurements and
from cosmological data.
\item
Are neutrinos Dirac or Majorana particles?
There is a strong experimental effort to detect
neutrinoless double-$\beta$ decay,
whose existence would imply that neutrinos are Majorana particles
(which is theoretically favored, especially in view of the see-saw mechanism).
\item
Are the
$\bar\nu_\mu\to\bar\nu_e$
transitions observed in the short-baseline LSND experiment~\cite{LSND} real?
The answer will hopefully come soon from the MiniBooNE experiment~\cite{hep-ex/0210020}.
\item
Which is the number of massive neutrinos?
If it is larger than three there may be transitions of active neutrinos into
sterile states in oscillation experiments.
\item
If the three-neutrino mixing scheme will be confirmed,
is it normal or inverted?
These two schemes may be distinguished in future
long-baseline experiments
through different matter effects if $|U_{e3}|$ is not too small.
Eventually,
the two schemes may be distinguished
by a direct measurement of all three masses.
\item
Is there CP violation in the lepton sector?
The possibility to measure CP violation in future long-baseline experiments
is currently under study
(see the contribution of M. Mezzetto in these Proceedings).
\end{enumerate}
In conclusion,
I think that the future promises to be very interesting for neutrino physics.

\section*{References}

\end{document}